# Crystallography and Chemistry of Perovskites

Mats Johnsson [a] and Peter Lemmens [b]

a   Dept. Inorg. Chemistry, Stockholm University, S-106 91 Stockholm, Sweden

b   Inst. Phys. for Condensed Matter, TU Braunschweig, D-38106 Braunschweig, Germany

**Abstract**

Despite the simplicity of the original perovskite crystal structure, this family of compounds shows an enormous variety of structural modifications and variants. In the following, we will describe several examples of perovskites, their structural variants and discuss the implications of distortions and non-stoichiometry on their electronic and magnetic properties.

**Introduction**

The structural family of perovskites is a large family of compounds having crystal structures related to the mineral perovskite $CaTiO_3$. In the ideal form the crystal structure of cubic $ABX_3$ perovskite can be described as consisting of corner sharing $[BX_6]$ octahedra with the A cation occupying the 12-fold coordination site formed in the middle of the cube of eight such octahedra. The ideal cubic perovskite structure is not very common and also the mineral perovskite itself is slightly distorted. The perovskite family of oxides is probably the best studied family of oxides. The interest in compounds belonging to this family of crystal structures arise in the large and ever surprising variety of properties exhibited and the flexibility to accommodate almost all of the elements in the periodic system. Pioneering structural work on perovskites were conducted by Goldschmidt *et al* in the 1920:s that formed the basis for further exploration of the perovskite family of compounds [1]. Distorted perovskites have reduced symmetry, which is important for their magnetic and electric properties. Due to these properties, perovskites have great industrial importance, especially the ferroelectric tetragonal form of $BaTiO_3$.

**The crystal structure of perovskite**

If the large oxide ion is combined with a metal ion having a small radius the resulting crystal structure can be looked upon as close packed oxygen ions with metal ions in the interstitials. This is observed for many compounds with oxygen ions and transition metals of valence +2, *e.g.* NiO, CoO, and MnO. In these crystal structures the oxygen ions form a cubic close packed lattice (ccp) with the metal ion in octahedral interstitials (*i.e.* the rock salt structure). Replacing one fourth of the oxygen with a cation of approximately the same radius as oxygen (*e.g.* alkali, alkali earth or rare earth element) reduces the number of octahedral voids, occupied by a small cation, to one fourth. The chemical formula can be written as $ABX_3$ and the crystal structure is called perovskite. X is often oxygen but also other large ions such as $F^-$ and $Cl^-$ are possible.

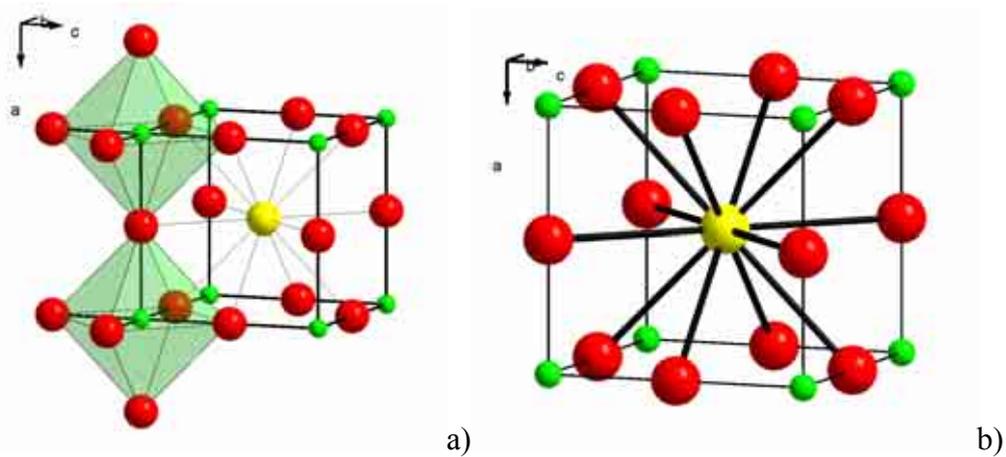

**Figure 1**   Outline of the ideal cubic perovskite structure SrTiO$_3$ that has **(a)** a three dimensional net of corner sharing [TiO$_6$] octahedra with **(b)** Sr$^{2+}$ ions in the twelve fold cavities in between the polyhedra.

The idealized cubic structure is realized, *e.g.* in CaRbF$_3$ and SrTiO$_3$. The latter can be described as Sr$^{2+}$ and O$^{2-}$ ions forming a cubic close packed lattice with Ti$^{4+}$ ions occupying the octahedral holes created by the oxygens. The perovskite structure has a three dimensional net of corner sharing [TiO$_6$] octahedra with Sr$^{2+}$ ions in the twelve fold cavities in between the polyhedra, see Figure 1. In the cubic ABX$_3$ perovskite structure ($a$ = 3.905 Å, space group Pm-3m, Z = 1) the A atoms are in Wyckoff position 1b, ½,½,½; the B atoms in 1a, 0,0,0; and the X atoms in 3d ½,0,0; 0,½,0; 0,0,½, all special positions. If the position of the Sr$^{2+}$ ion (A) is vacant the remaining framework is that of the ReO$_3$ type. Partial occupation of the A position occurs in the cubic tungsten bronzes A$_x$WO$_3$ (A = alkali metal, 0.3 ≤ x ≤ 0.93). The ReO$_3$ structure type can be converted to a more dense packing by rotating the octahedra until a hexagonal close packing is obtained of the RhF$_3$ type. The void in the centre has then an octahedral surrounding. If this octahedral hole is occupied we have the ilmenite structure, FeTiO$_3$. The perovskite structure is known to be very flexible and the A and B ions can be varied leading to the large number of known compounds with perovskite or related structures. Most perovskites are distorted and do not have the ideal cubic structure.

Three main factors are identified as being responsible for the distortion: Size effects, deviations form the ideal composition and the Jahn-Teller effect. It is rare that a distortion of a certain perovskite compound can be assigned to a single effect. In most cases several factors act on the structure. As an example of the complexity BaTiO$_3$ has four phase transitions on heating: rhombohedral (R3m) — –90 °C → orthorhombic (Amm2) — 5 °C → tetragonal (P4mm) — 120 °C → cubic (Pm-3m).

1. Size effects

In the ideal cubic case the cell axis, $a$, is geometrically related to the ionic radii ($r_A$, $r_B$, and $r_O$) as described in equation (1):

$$a = \sqrt{2}(r_A + r_O) = 2(r_B + r_O) \qquad (1)$$

The ratio of the two expressions for the cell length is called the Goldschmidt's *tolerance factor t* and allows us to estimate the degree of distortion. It is based on ionic radii *i.e.* purely ionic bonding is assumed, but can be regarded as an indication for compounds with a high degree of ionic bonding; it is described in equation (2).

$$t = \frac{(r_A + r_O)}{\sqrt{2}(r_B + r_O)} \qquad (2)$$

The ideal cubic perovskite SrTiO$_3$ has $t = 1.00$, $r_A = 1.44$ Å, $r_B = 0.605$ Å, and $r_O = 1.40$ Å. If the A ion is smaller than the ideal value then $t$ becomes smaller than 1. As a result the [BO$_6$] octahedra will tilt in order to fill space. However, the cubic structure occurs if $0.89 < t < 1$ [2-3]. Lower values of t will lower the symmetry of the crystal structure. For example GdFeO$_3$ [4] with $t = 0.81$ is orthorhombic ($r_A = 1.107$ Å and $r_B = 0.78$ Å), see Figure 2a. Also the mineral perovskite itself, CaTiO$_3$, has this structure. With values less than 0.8 the ilmenite structure is more stable.

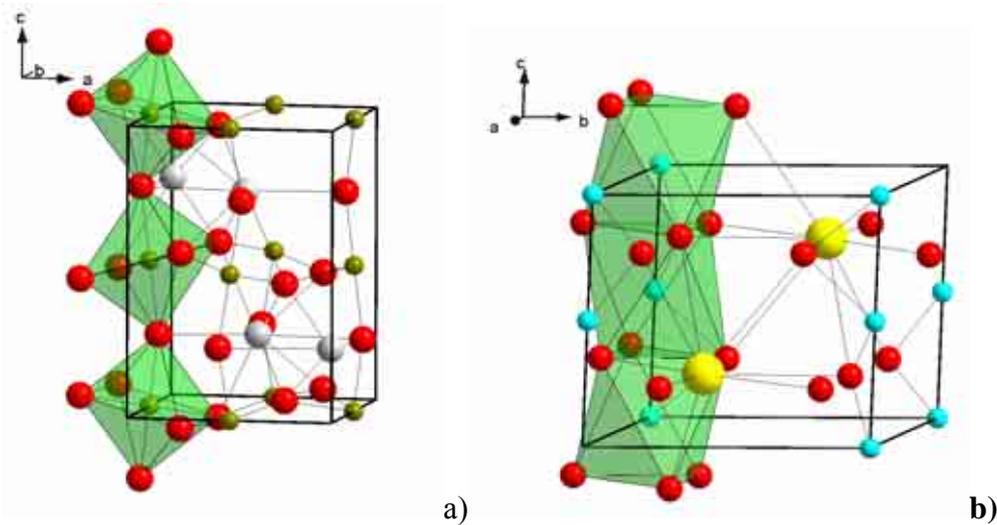

a)          b)

**Figure 2**    **(a)** Low values of the tolerance factor $t$ will lower the symmetry of the crystal structure. GdFeO$_3$ with $t = 0.81$ have tilted [FeO$_6$] octahedra and crystallize in the orthorhombic system ($r_A = 1.107$ Å and $r_B = 0.78$ Å). **(b)** If $t$ is larger than 1 due to a large A or a small B ion then hexagonal variants form of the perovskite structure. The $t$ value for BaNiO$_3$ is 1.13 ($r_A = 1.61$ Å and $r_B = 0.48$ Å).

On the other hand if $t$ is larger than 1 due to a large A or a small B ion then hexagonal variants of the perovskite structure are stable, *e.g.* BaNiO$_3$ type structures. In this case the close packed layers are stacked in a hexagonal manner in contrast to the cubic one found for SrTiO$_3$, leading to face sharing of the [NiO$_6$] octahedra, see Figure 2b. The $t$ value for BaNiO$_3$ is 1.13 ($r_A = 1.61$ Å and $r_B = 0.48$ Å). Since perovskites are not truly ionic compounds and since the $t$ values also depend on what values are taken for the ionic radii, the tolerance factor is only a rough estimate.

2. Changing the composition from the ideal ABO$_3$

An example is the family of compounds SrFeO$_x$ ($2.5 \leq x \leq 3$). The valency of the Fe ions can be changed by heating a sample in either an oxidizing or a reducing environment. As a result the oxygen content can vary in between 2.5 and 3. In for example SrFeO$_{2.875}$ some Fe ions can be assigned to the oxidation state +3 and others to +4. The oxygen vacancies order so that FeO$_5$ square pyramids are formed, see Figure 3. The SrFeO$_x$ compounds are examples of defect perovskites. Their chemistry can be described according to the homologous series A$_n$B$_n$O$_{3n-1}$, n = 2–∞. Several other types of vacancy orderings are known, *e.g.* the structures of Ca$_2$Mn$_2$O$_5$ and La$_2$Ni$_2$O$_5$ having n = 2 are shown in Figure 4a-b.

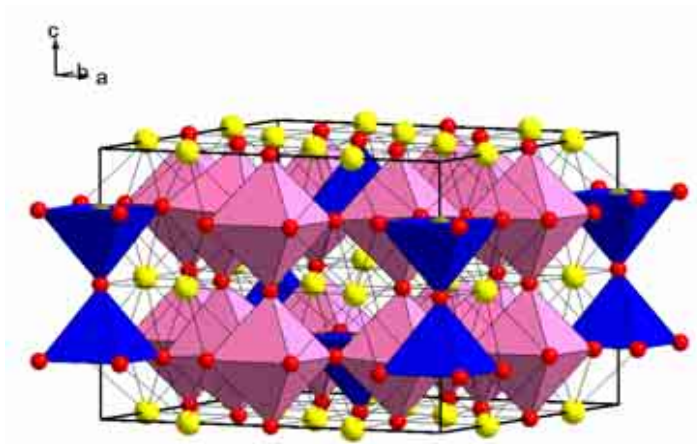

**Figure 3** Ordering of oxygen vacancies in $SrFeO_{2.875}$ (= $Sr_8Fe_8O_{23}$). Fe ions are located in both square pyramids and in octahedra.

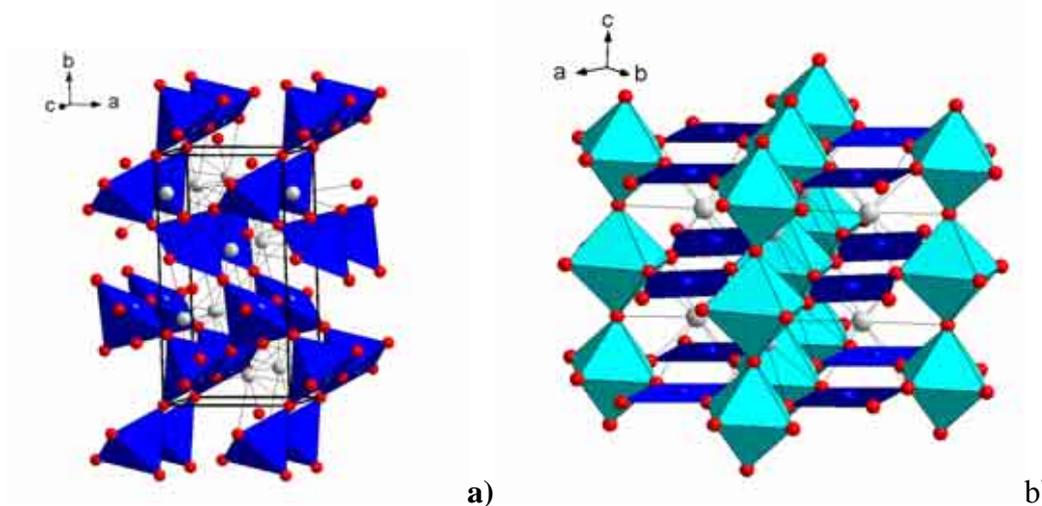

**a)** **b)**

**Figure 4** Ordering of oxygen vacancies in **(a)** $Ca_2Mn_2O_5$ having [$MnO_5$] square pyramides and **(b)** $La_2Ni_2O_5$ having [$NiO_6$] octahedra and [$NiO_4$] square planes.

3. Jahn – Teller effects

In some perovskites the distortion of the structure can be assigned to Jahn – Teller active ions at the B position. For example in $LnMnO_3$ (Ln = La, Pr or Nb) with $Mn^{3+}$ ions the $3d^4$ electrons divide up into 3 $t_g$ and 1 $e_g$ electron. The odd number of electrons in the $e_g$ orbital causes an elongation of the [$MnO_6$] octahedron.

**Superstructures related to the perovskite structure**

If we double all three unit cell edges of the cubic perovskite structure it is possible to occupy equivalent positions with atoms of different elements, see Figure 5. In $K_2NaAlF_6$ the $K^+$ and the $F^-$ ions take the $Ca^{2+}$ and the $O^{2-}$ positions respectively of the perovskite. The one-to-one relation can be recognized by comparing with the doubled formula of perovskite. The comparison also shows how the octahedral $Ti^{4+}$ position shift into two sites for $Na^+$ and $Al^{3+}$ [3]. In kryolite, $Na_3AlF_6$, the $Na^+$ ions occupy two different positions, namely the $Sr^{2+}$ and the $Ti^{4+}$ positions of the doubled perovskite cell, *i.e.* positions with coordination numbers of 6 and 12. Since this is not convenient for ions of the same size, the structure experiences some distortion.

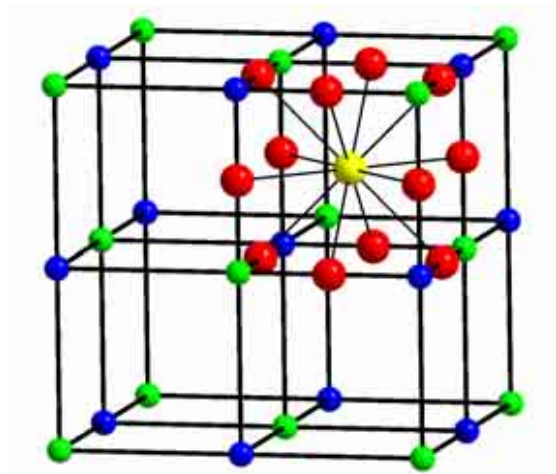

**Figure 5**  Superstructures of the perovskite type. If the unit cell edges are doubled it is possible to occupy equivalent positions with atoms of different elements. The one-to-one relation can be recognized by comparing with the doubled formula of perovskite.

| Structure type | Example | Yellow | Red | Green | Blue |
|---|---|---|---|---|---|
| Perovskite | $SrTiO_3$ | $Sr^{2+}$ | $O^{2-}$ | $Ti^{4+}$ | $Ti^{4+}$ |
| Elpasolite | $K_2NaAlF_6$ | $K^+$ | $F^-$ | $Na^+$ | $Al^{3+}$ |
| Kryolite | $(NH_4)_3AlF_6$ | $NH_4^+$ | $F^-$ | $NH_4^+$ | $Al^{3+}$ |
| $K_2PtCl_6$ | | $K^+$ | $Cl^-$ | – | $Pt^{4+}$ |

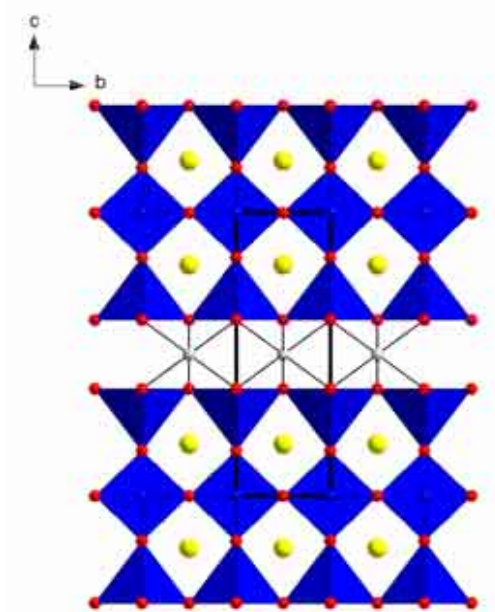

**Figure 6**  The high temperature superconductor $YBa_2Cu_3O_{6.96}$. The structure is a superstructure of perovskite with approximately 2/9 of the oxygen positions vacant, in such way that 2/3 of the Cu atoms have $[CuO_5]$ square pyramidal coordination and 1/3 have square-planar $[CuO_4]$ coordination. The perovskite structure is attained by inserting oxygen atoms in between the yttrium atoms (grey) and in between the $[CuO_4]$ square planes.

Perovskites of the type $ACuO_{3-\delta}$ which have Cu atoms in the octahedral sites are deficient in oxygen, Alkaline earth and trivalent ions ($Y^{3+}$, lanthanoids, $Bi^{3+}$, $Tl^{3+}$) occupy the A site. A typical composition is $YBa_2Cu_3O_{7-x}$ with x ≈ 0.04. These compounds are high temperature superconductors. The structure is a superstructure of perovskite, but with approximately 2/9 of the oxygen positions vacant, in such way that 2/3 of the Cu atoms have square pyramidal coordination and 1/3 have square-planar coordination, see Figure 6. The cobaltite $GdBaCo_2O_{5.5}$ is another example of an oxygen deficient perovskite related structure where the $Co^{3+}$ ions have octahedral and square pyramidal coordination [5], see Figure 7.

Also the Brownmillerite structure is an oxygen deficient superstructure of cubic perovskite with an ordering of oxygen vacancies. Ruddlesden and Popper designed a series of homologous compounds with the general formula $AO(ABO_3)_n$ where AO represent a rock salt structure layer separating blocks of perovskite layers characterized by n = 1,2,3…∞. Examples are the high $T_c$ superconductor prototype $La_2CuO_4$ and the 2D quantum antiferromagnet $La_2NiO_4$.

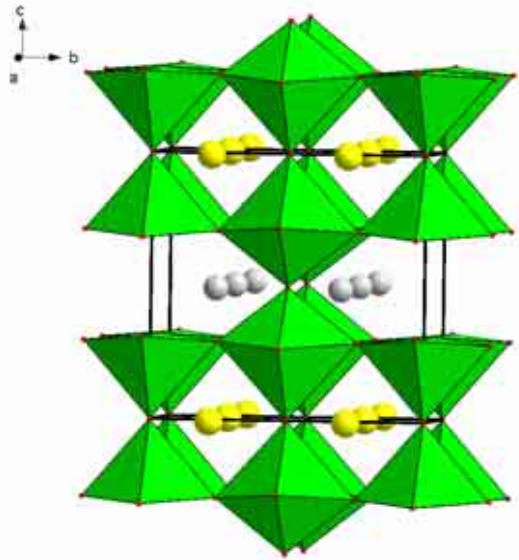

**Figure 7** The cobaltite $GdBaCo_2O_{5.5}$ has a perovskite related structure were 1/12 of the oxygen atoms are missing leading to that 50% of the Co atoms have square pyramidal $[CoO_5]$ coordination and 50% of the Co atoms have octahedral $[CoO_6]$ coordination. Ba atoms are yellow and Gd atoms are grey.

**Electronic and magnetic properties of perovskites**

Perovskites with transition metal ions (TMI) on the B site show an enormous variety of intriguing electronic or magnetic properties. This variety is not only related to their chemical flexibility, but also and to a larger extent related to the complex character that transition metal ions play in certain coordinations with oxygen or halides [6]. While magnetism and electronic correlations are usually related to unfilled 3d electron shells of the TMI, pronounced dielectric properties are connected with filled 3d electron shells. Multiferrocity, a coexistence of spontaneous ferroelectric and ferromagnetic moments, is a rare phenomenon due to the small number of low-symmetry magnetic point groups that allow a spontaneous polarization [7]. Nevertheless, in the presence of competing interactions [8], canted moments [9-10] or in composites [11] large magneto-capacitive couplings have been reported.

In the following we will discuss examples of material properties in which transition metal perovskites and related structures prove to be outstanding. To some extend these aspects also touch application areas, as *e.g.* capacitors, transducers, actuators, sensors and electrooptical switches.

*Dielectric and ferroelectric perovskites*

High dielectric permittivity ($\varepsilon$) or ferroelectric materials are of enormous importance as electroceramics for engineering and electronics. Perovskites, *e.g.* titanium or niobium perovskites, $BaTiO_3$ and $LiNbO_3$, have been intensively studied in the past [12]. A large $\varepsilon$ is based on collective polar displacements of the metal ions with respect to the oxygen sublattice and is a highly nonlinear and anisotropic phenomenon. The phase transition that leads to ferroelectricity is usually described by a soft-mode model [13].

To optimize dielectric and mechanical properties several routes have been followed from the structurally simple $BaTiO_3$ via the solid solution system $Pb(Zr,Ti)O_3$ to other distinct families of materials. These routes explicitly take into account the flexibility for chemical manipulation and "docility" of the perovskites [12]. One of them is the relaxor ferroelectric. It is genuinely based on a multi-element substituted Pb titanate ($PbTiO_3$) with the composition $A(B'B'')O_3$ with a random occupation of the A and B sites by metal ions of different valence.

Relaxor ferroelectrics show enormously large dielectric constants, a pronounced frequency dispersion and variation of $\varepsilon$ as function of temperature. These effects are due to slow relaxation processes for temperatures above a glass transition [14]. The length scales of fluctuating composition and spontaneous polarization are 2-5 nm, i.e. the effects are based on electronic inhomogeneities and the existence of polar nanoregions. The lattice part of the response is considered to be a local softening of transverse-optical phonon branch that prevents the propagation of long-wavelength ($q = 0$) phonons. It is interesting to note that the fundamental limit, the superparaelectric state, is still not reached for such small length scales [15]. Generic examples for relaxor ferroelectrics are PZT: $Pb(Zn_{1/3}Nb_{2/3})O_{3-x}PbTiO_3$ and PMN: $Pb(Mg_{1/3}Nb_{2/3})O_{3-x}PbTiO_3$, with PZT having a higher temperature scale compared to PMN.

Incipient ferroelectrics or quantum paraelectrics can be regarded as almost ferroelectric crystals [16]. Examples are $KTaO_3$ and $SrTiO_3$ [17]. Pronounced quantum fluctuations of ions suppress the phase-transition into the ferroelectric state and stabilize the soft transverse optical mode. The dielectric susceptibility shows a divergence in the limit T to 0 K together with pronounced phonon anharmonicities [17]. In these systems even minor substitutions or doping can induce phase transitions into ferroelectric states. Finally, we mention perovskite related oxides with giant dielectric constants (GDC) where no evidence for a ferroelectric instability exists. These nonintrinsic permittivities are attributed to barrier layers and surface effects [18]. Examples are $CaCu_3Ti_4O_{12}$, [19] and the Li-ion conductor material $La_{0.67}Li_{0.25}Ti_{0.75}Al_{0.25}O_3$ [20].

*Magnetism and electronic correlations*

Magnetism or orbital (electronic) ordering phenomena of various kinds are observed in perovskites with TMI that have unfilled 3d electron shells. Electronic correlations [21] of such 3d states are generally strong, as the ratio $U_d/W$ of the Coulomb repulsion energy $U_d$ vs. the bandwidth W is larger compared to other electronic states, *i.e.* they have a more local character and a tendency for insulating states or metal-insulator transitions [22]. Hopping and superexchange of these electrons takes place via oxygen sites due to the overlap of the respective wave function. Thereby, the properties and phase diagrams of a perovskite strongly depend on nonstoichiometries and even more on tilting or distortions of the [$BO_6$] octahedra. Further aspects rely on order/disorder processes of the orbital part of the 3d wave function, charge doping and charge/orbital inhomogeneous states that lead to colossal response, *e.g.* to external magnetic fields [23].

Before, however, considering such effects the properties of the system are given by a hierarchy of energies based on the electronic structure, *i.e.* the number of 3d electrons, the Hunds Rule coupling, the crystalline electric field or Jahn-Teller splitting of the 3d electron states and finally due to exchange energies.

*Cuprates, Jahn-Teller distortions and high temperature superconductors*

This hierarchy of energies is well established for cuprates with $Cu^{2+}$ in a $3d^9$ configuration. The hole in the d shell has $e_g$ $x^2-y^2$ symmetry and contributes to an $s = 1/2$ spin moment. The orbital moment is quenched due to the crystalline electric field of the surrounding oxygen. The $e_g$ electron is Jahn-Teller active, *i.e.* local or collective configurations with oxygen in distorted octahedra are energetically preferable. Extreme limits are pyramidal [$CuO_5$] or even a planar [$CuO_4$] configuration of the oxygen neighbors. Thereby the superexchange and magnetic interactions between the $s = 1/2$ spin moments are restricted to a plane or, if building blocks are shifted by half a unit cell within the plane, to a quasi-one dimensional path. There are numerous realizations of such low-dimensional magnetic systems as in $Sr_2CuO_3$ (spin chain system) or $SrCu_2O_3$ (spin ladder system) [24]. Due to the small coordination number of the spin moments in one dimension and pronounced quantum fluctuations related to the small magnitude of the spin, such compound do generally not show long range ordering. Strong fluctuations are evident as broad maxima in the magnetic susceptibility and continua in inelastic neutron scattering.

Superexchange and electronic correlations restricted to a two dimensional, weakly doped plane are the key ingredients of high temperature superconductors. The crystal structure of the prototype and perovskite related compound $YBa_2Cu_3O_{7-x}$ is shown in Figure 6. For $x \approx 1$ the resulting $Cu^{2+}$ with $s = 1/2$ moments show long antiferromagnetic range ordering with a Néel temperature of more than 500 K. This high ordering temperature marks the exceptionally large energy scales and strong correlations involved in these materials. With smaller x doping is induced that leads to a drastic drop of the Néel temperature and the onset of high temperature superconductivity. The maximum superconducting transition temperature is $T_{cmax}$= 92 K for this system.

Electronic correlations are essential to understand the effect of doping. The electronic structure of cuprates in the vicinity of the Fermi level is given by an occupied low-energy and an unoccupied, high energy band, the lower and the upper Hubbard band, separated by the Coulomb repulsion energy $U_d$ of the 3d electrons. High temperature superconductors are charge transfer insulators, *i.e.* the oxygen is included in this scheme as an occupied, non-bonding 2p band separated by a smaller charge transfer energy $\Delta$ from the upper Hubbard band ($\Delta < U_d$). The doping process consists of introducing a novel correlated electron state, the Zhang-Rice singlet state [22], in the proximity of the oxygen band. This state of hybridized Cu and O character leads to a transformation from a long range Néel state to a high temperature superconductor. Although the number of known high temperature superconductors seam to be large (of the order of 20 compounds) they all rely on this scheme of a doped, two dimensional perovskite related structure with pronounced electronic correlations [22, 25-26].

*Cobaltates, spin state transitions and oxygen deficiency*

If the above mentioned hierarchy of energies is not well defined, the compound chooses certain ways to lift degeneracies of the electronic system. Important are spin state transitions or crossover behaviour, a partial metallization of 3d electrons, or charge disproportionation of the TMI sites. Perovskites based on cobalt and vanadium serve as model systems for such effects and the resulting interplay of electronic and structural degrees of freedom. In the following we will discuss briefly two cobaltates to give an example for the resulting complexities.

In the cobalt perovskite $LaCoO_3$ with the same crystal structure as is shown in Figure 1 all three spin states of $Co^{3+}$ ($3d^6$) are close to degenerate. As these states correspond to slightly different ionic radii, with decreasing temperatures a crossover of the dominant populations from high spin (s = 2), intermediate spin (s = 1) to low spin (s = 0) $Co^{3+}$ $3d^6$ states takes place. This process is mainly controlled by temperature and has no evident collective character. The magnetic susceptibility shows a broad maximum and a strong decrease at low temperatures [27]. The different ionic radii of the spin states also couple the electronic configurations of the TMIs only weakly to other properties of the compound.

An ordered oxygen deficiency leads to a multiplication of the unit cell volume. It has also profound influence on the electronic and magnetic properties of the compound [28]. Due to the smaller coordination number of some TMI sites the respective bandwidth is reduced and with increasing electronic correlations the tendencies for charge/orbital ordered states is enhanced. In Figure 7 the perovskite cobaltite $GdBaCo_2O_{5.5}$ is depicted. Oxygen defects form chains of $[CoO_5]$ pyramids and $[CoO_6]$ octahedra along the crystallographic *a*-axis. Compared to the ideal perovskite $LaCoO_3$, the behaviour is rather complex and highly collective. The phase diagram contains a metal-insulator transition and three different magnetic phases that include spin state ordering [29-30].

*Manganites and orbital degrees of freedom*

In the manganite $(La,Sr)MnO_3$ the ratio $La^{3+}/Sr^{2+}$ determines the oxidation state of Mn and thus the ratio $Mn^{3+}/Mn^{4+}$. This corresponds to the number of Mn sites with a single 3d $e_g$ orbital occupied. Double exchange describes the situation where these states simultaneously hop via $Mn^{4+}$ ions. The remaining 3 $t_{2g}$ electrons on each Mn ion sum up to s=3/2 due to Hunds rule coupling and form a 'rigid background'. The band width and charge transport solely given by $e_g$ states dependents on the spin and orbital orientation of the exchange partners.

As function of composition different magnetic ground states and orbitally/charge ordered structures are observed. These degrees of freedom react rather cooperative due to the strong interlink of the octahedra in the perovskite structure [31]. Pronounced effects are observed in all physical quantities. However, most spectacular is the colossal magneto-resistance (CMR) at the borderline between a ferromagnetic insulating and ferromagnetic metallic phase. For a more complete treatment of this increasingly rich field of research including a discussion of relevant vanadium and titanium perovskites we refer to reviews [22-23, 32-34] and recent focus issues [35] of international journals.

**Synthesis**

Many perovskites are synthesised by solid state reactions giving polycrystalline samples. The starting materials are then usually simple binary oxides or pure elements reacted at relatively high temperatures. This synthesis technique involves problems due to the fact that certain starting oxides (*e.g.* PbO) may vaporize. The reaction temperature can be lowered by applying microwave synthesis techniques and thus minimizing the loss of volatile starting components. Hydrothermal synthesis techniques have been applied to manufacture nanopowders of *e.g.* BaTiO$_3$. Powders and thin films with controlled levels of dopants have been prepared with the Sol-Gel technique using metal alkoxides as precursors. Thin films of ferroelectrics have been successfully prepared by physical vapor deposition (PVD) or pulsed laser deposition (PLD).

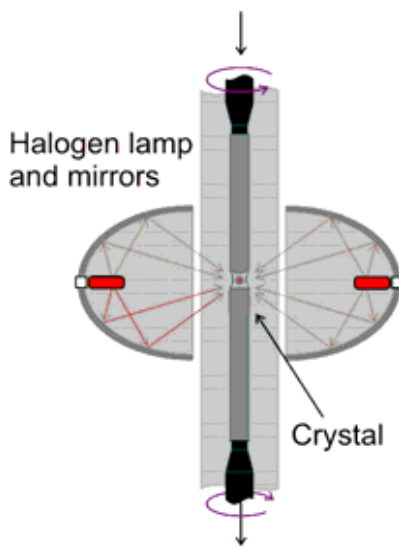

**Figure 8**  Outline of the principles for a floating zone mirror furnace. The crystal is grown by moving sintered powder rods through an optically heated floating zone.

During recent years, several research groups have succeeded in growing single crystals of several families of perovskite related compounds from molten alkali carbonates or other fluxes such as hydroxides or halides. Large single crystals (>10 cm long) of manganites and other oxides have been grown by utilizing optically heated floating zone techniques with oxygen injected into the furnace around the molten zone, see Figure 8. The review [1] and references therein is referred to for further aspects of synthesis techniques.

**Acknowledgement:** This work has been carried out with financial support from the Swedish Research Council, INTAS and by the DFG through SPP 1073. We acknowledge discussion with Yurii Pashkevich and Rainer Waser.